\documentclass[aps,prd,onecolumn,nofootinbib,superscriptaddress]{revtex4}
\pdfoutput=1
\usepackage{graphicx}
\usepackage{amsmath}
\usepackage{amsfonts}
\usepackage{amssymb,ulem}
\usepackage{color}%
\usepackage{dcolumn}

\usepackage{MnSymbol,wasysym}
\usepackage{braket}
\usepackage{eurosym}
\usepackage{subfigure}
\usepackage{multirow}
\usepackage{float}
\usepackage[usenames,dvipsnames,svgnames]{xcolor}

\newcommand{\RNum}[1]{\uppercase\expandafter{\romannumeral #1\relax}}
\usepackage[colorlinks=true,linkcolor=blue,urlcolor=blue,filecolor=black,citecolor=red,pdfstartview=FitV,pdftitle={},pdfsubject={},pdfkeywords={},pdfpagemode=None,bookmarksopen=true]{hyperref}
\usepackage[title]{appendix}
\makeatletter
\newcommand*{\rom}[1]{\expandafter\@slowromancap\romannumeral #1@}
\makeatother

\begin{document}
\baselineskip=0.5 cm

\title{Using the shadow of a black hole to examine the energy exchange between axion matter
 and a rotating black hole}
\author{Xiao-Mei Kuang}
\email{xmeikuang@yzu.edu.cn (corresponding author)}
\affiliation{Center for Gravitation and Cosmology, College of Physical Science and Technology, Yangzhou University, Yangzhou, 225009, China}

\author{Yuan Meng}
\email{mengyuanphy@163.com}
\affiliation{Center for Gravitation and Cosmology, College of Physical Science and Technology, Yangzhou University, Yangzhou, 225009, China}

\author{Eleftherios Papantonopoulos}
\email{lpapa@central.ntua.gr}
\affiliation{Physics Division, School of Applied Mathematical and Physical Sciences, National Technical University of Athens, 15780 Zografou Campus,
    Athens, Greece.}

\author{Xi-Jing Wang}
\email{xijingwang01@163.com (corresponding author)}
\affiliation{Department of Astronomy, School of Physics and Technology, Wuhan University, Wuhan 430072, China}

\begin{abstract}
We find that a \textit{slowly} rotating axion-modified black hole resulting from the backreaction of an axion field on a rotating Kerr black hole  can have a \textit{D-shaped} shadow as that for a \textit{highly} counter-rotating Kerr black hole. This attributes to the fact that the energy exchange between the axion matter and the black hole influences the rotation of the black hole, so the black hole angular momentum first decreases to zero and then starts to rotate to the opposite direction. Further increasing the coupling leads to  \textit{``human-face-like" shaped} shadows and new lensing due to the chaotic scattering, which are novel and drastically different from those for Kerr black hole. Our analysis provides the first counterexample to that slowly rotating black hole has nearly circular shadow.
\end{abstract}

\maketitle
\section{Introduction}
In a realistic Universe, the black holes (BHs) are rotating which are described by the Kerr metric~\cite{Kerr:1963ud}. For a rotating BH described by the Kerr metric, it is more complicated to describe the behaviour of matter than in the non-rotating case. The metric in this case is  only stationary,  and the spherical symmetry has been replaced by axial symmetry. A further difficulty is the complexity of the coordinate systems for figuring out various processes
near a Kerr BH. Till now, the behaviour of matter in the vicinity of rotating BHs has been extensively studied.
In the pioneering work \cite{Bardeen:1972fi}, the authors have studied the physical processes around the BHs, as well as their properties and interactions with the surrounding matters. In this scenario, one usually assumes that the background metric is slowly rotating to avoid the complexity in the perturbation analysis, the essence of which is to assume that the non-rotating matter  cancels out the ``frame-dragging” effects of the BH's rotation. In a sense, such a matter is known to co-rotate with the BH such that the physical processes can be analyzed in its own frame.  The authors of  \cite{Bardeen:1972fi}  used the method of locally non-rotating frame, which could provide a clear picture to illustrate the process of energy extraction.  However, the amount of such energy for realistic BHs is insignificant for astrophysics.

The modified {theory} of gravity like the Chern-Simons (CS)-gravity theory takes the action~\cite{Jackiw:2003pm, Zanelli:2005sa,Alexander:2009tp,Yunes:2009hc}
\begin{eqnarray}
\begin{split}
S
=\int d^4x\sqrt{-g}\left[\frac{R}{2\kappa^2}-\frac{1}{2}(\partial_{\mu}b)(\partial^\mu b)-A b R_{CS}\right],
\end{split}
\end{eqnarray}
where $b$ is a pseudoscalar denoting the axion matter field, $A$ is the coupling parameter of the axion field to the gravitational CS term $R_{CS}$ with length dimension, and $\kappa^2=8\pi G$ with $G$ the Newton's gravitational constant in four dimensional spacetime. Here, we shall use natural units $G = c = 1$. It is noticed that the CS modification is a total derivative, and thus innocuous unless the axion-like pseudoscalar fields couple to it. {The coupling gives a scalar field equation sourced by the Pontryagin density, while modifies  the metric field equations into third-order derivatives, which may cause the full CS gravity not be a well-posted initial value problem \cite{Delsate:2014hba} such that stable BH solutions are difficult to be found. But CS gravity is usually treated as an
effective field theory characterising the low-energy-limit behaviors of string theory,  thus, it is always treated to perturbatively deform from {General Relativity (GR)}, reducing the order of the field equations \cite{Campbell:1990ai,Campbell:1990fu,Duncan:1992vz}.} In this case, the interaction could deform non-trivially the gravitational field of the spinning BHs, leading to a modified Kerr geometry (Kerr-like BH solution), which contains second-order curvature corrections.
In string theory framework,  the axion fields are part of the excitation spectrum, which in general have two types~\cite{Svrcek:2006yi}. One type  is the so-called string-model independent axion. In (3+1)-dimensional case, this axion is dual~\cite{Duncan:1992vz,Svrcek:2006yi} to the field strength of the Kalb-Ramond, spin-one, antisymmetric-tensor field, which plays the role of a totally antisymmetric torsion in the geometry. The other type of axion arises from the compactification of the extra dimensions in string theory.
Potential signatures of CS theories in the observational gravitational waves (GW),  emitted from the inspiral process of stellar compact objects into the massive BHs, with both intermediate and extreme mass ratios, have been explored  in \cite{Sopuerta:2009iy}. {T}he authors of \cite{Canizares:2012is} made a first attempt to answer the question that to some extent the observations from extreme mass ratio inspirals with a space-based gravitational wave observatory such as LISA  might have potential to distinguish CS theories from {GR}. {By simulating GW150914 with numerically binary {BH} signals in a CS theory, the leading order CS corrections to the merger waveform as well as the ringdown waveform and quasi-normal mode spectrum have been present in \cite{Okounkova:2019zjf}. Furthermore, the possibility of detecting CS amplitude birefringence with GW detectors have been discussed in \cite{Yunes:2010yf,Yagi:2017zhb,Nojiri:2019nar, Zhao:2019szi}, which later been verified with real GW data from binary {BH} mergers \cite{Okounkova:2021xjv}.  }

In such CS theories, rotating BHs have been extensively studied. First, in~\cite{Campbell:1990ai,Campbell:1990fu,Duncan:1992vz} and  recently in  \cite{Boulware:1985wk,Gonzalez:2010vv,Yagi:2012ya}, they were found as solutions to second-order field equations for the metric stemming from string theories with axions.
{Afterwards, focusing on some slowly rotating hairy BHs in this gravity, the BH perturbations \cite{Wagle:2023fwl}, stable quasinormal spectrums \cite{Wagle:2021tam,Srivastava:2021imr} and superradiant instability \cite{Alexander:2022avt}  have been extensively explored.}
In particular, in \cite{Chatzifotis:2022mob}, the authors extended the non-trivially previous results~\cite{Campbell:1990ai,Campbell:1990fu,Duncan:1992vz,Yunes:2009hc,Myung:2020etf} and investigated the slowly rotating BH solutions, in the presence of axion fields  back reacting on the geometry, of which the metric is
\begin{eqnarray}
\begin{split}
ds^2=-\left(1-\frac{2M}{r}\right)dt^2+\frac{1}{\left(1-\frac{2M}{r}\right)}dr^2+r^2\left(d\theta^2+\sin^2{\theta}d\phi^2\right)-2r^2a \sin^2\theta W(r)dtd\phi~, \label{eq-metric}
\end{split}
\end{eqnarray}
with
\begin{eqnarray}
\begin{split}
W(r)=\frac{2M}{r^3}-\frac{\gamma^2(189 M^2+120M r+70r^2)}{14r^8}+\mathcal{O}(A^{2n})~,  ~~~~ \gamma^2=\frac{A^2 \kappa^2}{M^4}~,
\label{eq-metric-1}
\end{split}
\end{eqnarray}
where $M$, $\gamma$ and $a$ are the mass of BH, coupling parameter, and rotation parameter, respectively.
The back reaction of the {axion} field on the slowly rotating  metric has been studied to arbitrary order  in the perturbation theory with an appropriate parameter, $\gamma$, which is proportional to the axion-CS coupling strength. It is explicit that instead of the coupling constant itself,  the dimensionless parameter $\gamma$ appears in the correction function of the modified Kerr-like BH solution and measures the strength of the back reaction on the geometry.
The expansion generated by this perturbation theory allows one to  arbitrarily approach the horizon of the hairy BH. Thus, one could investigate the axionic matter's behaviour around the horizon of the slowly rotating Kerr-like metric and check how the matter distribution would affect the geometry itself. Especially, it also means that the axion-hair is not short and can extend into the innermost light ring, which is crucial to the observations from BH shadow and gravitational lensing aspects \cite{Ghosh:2023kge}.
It is known that the total angular momentum for Kerr BH is given by $Ma$ \cite{Carter:1968rr}. However, here the presence of the axion field coupled to the CS term, which back reacts on the original Kerr background,  implies that the total angular momentum of the sector should consist of the contributions from both the axion field and the hairy Kerr-like BH itself, which arises because of the axionic matters back reaction to the geometry \cite{Chatzifotis:2022mob,Chatzifotis:2022ene}.

Later, the study on this axion-modified-Kerr BH system was extended by seeking for potentially observable effects from the bound trajectories of satellites in such backgrounds \cite{Chatzifotis:2022ene}, in which the aforementioned perturbative parameter $\gamma$ plays a key role. They found that the axion matters surrounding the BH horizon acquires angular momenta in such a way that the total angular momentum of the sector consisting of BH and axionic matter remains constant $Ma$.
Consequently,  the BH angular momentum decreases as $\gamma$ increases, and at a critical value $\gamma_c$,  the BH starts to rotate to the opposite direction and the angular momentum increases when further increasing $\gamma$. Following the line, the angular momentum of the BH horizon can reach large values in magnitude, while the slowly rotating approximation is still valid with a total angular momentum {that is} equal to $Ma$, which was interpreted as the effects of a counterplay between two competing systems, i.e.,  the Kerr-like BH and the axionic matters rotating outside the BH horizon.

In this work, we study possible observational effects, like shadow and lensing \cite{Perlick:2021aok,Cunha:2018acu,Wang:2022kvg,Lukes-Gerakopoulos:2010ipp}, of  the  axion-modified-Kerr BH with the metric \eqref{eq-metric}. The coupling between the axion and gravity  modifies the equations of motions in such a way as to allow an exchange of energy between the axion field and gravity. We expect to probe the evidence  of energy exchange by BH shadow, and provide alternative BH shadow profile, since nowadays is golden era to study various phenomenological deviations from BH in GR but has potential to be detected by Event Horizon Telescope. Throughout the paper, all physical quantities are rescaled by $M$ so that we will work with $M=1$.

\section{Shadow and Lensing by Axion induced Kerr-like black hole}
\label{sec1}

In order to obtain the distorted apparent sky for a given observer, namely the shadow and lensing effect of Kerr-like BH, we shall compute photon geodesics in the above geometry \eqref{eq-metric}.
The equations of motion  of the photons can be determined from the Hamilton-Jacobi equation,
\begin{equation}
\mathcal{H}=-\frac{\partial S}{\partial\lambda}=\frac{1}{2}g^{\mu \nu}\frac{\partial S}{\partial x^\mu}\frac{\partial S}{\partial x^\nu}=0~,
\label{eq-Hamilton}
\end{equation}
where $\mathcal{H}$, $S$ and $\lambda$ are the canonical Hamiltonian, the Jacobi
action and affine parameter respectively. For the geometry, the system does not have the symmetry associated with Carter constant and cannot be integrated over it. Therefore, we use the numerical backward ray-tracing method to investigate the lensing effect and shadow of BH. To this end, we employ the setup in \cite{Bohn:2014xxa,Cunha:2015yba,Hu:2020usx,Zhong:2021mty}, so we consider that the BH is illuminated by the ball-like source which is placed at far distance, and the BH and the camera are inside the ball-like source. The observer uses the fish-eye camera to photograph the BH.

\begin{figure}[h]
\centering
\subfigure[\, Spherical source]
{\includegraphics[width=4.cm]{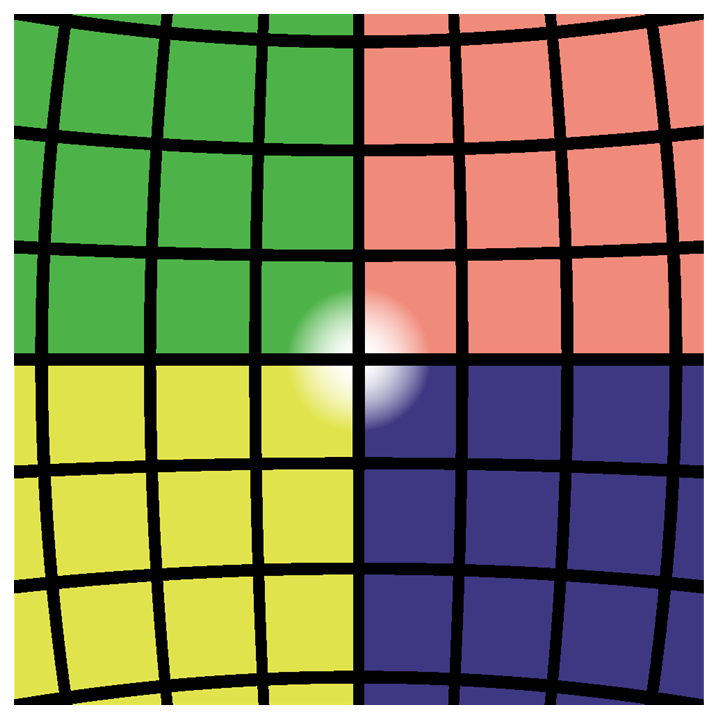}\label{fig:source1}\hspace{1mm}}
\subfigure[\, Schwarzschild]
{\includegraphics[width=4.cm]{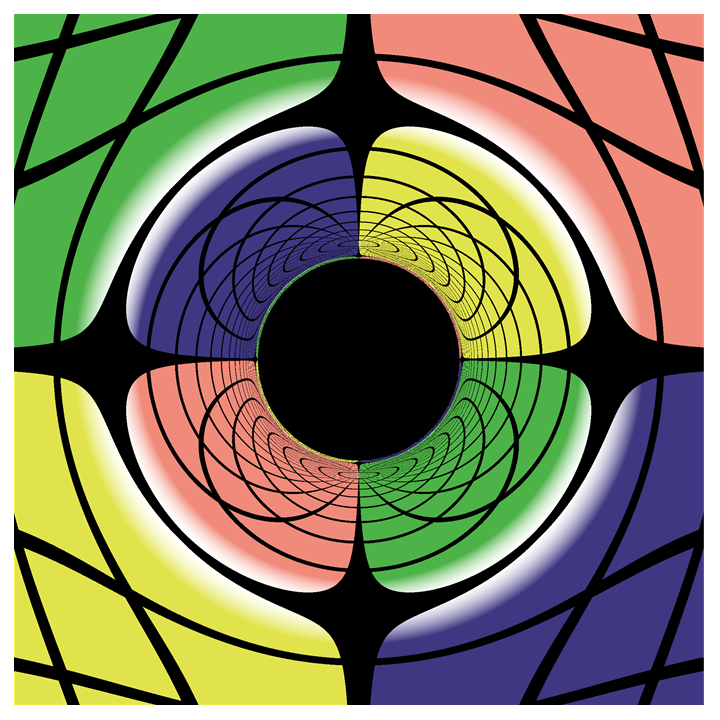}\label{fig:lensing1}}\hspace{1mm}
\subfigure[\, Kerr, $a=0.999$]
{\includegraphics[width=4.cm]{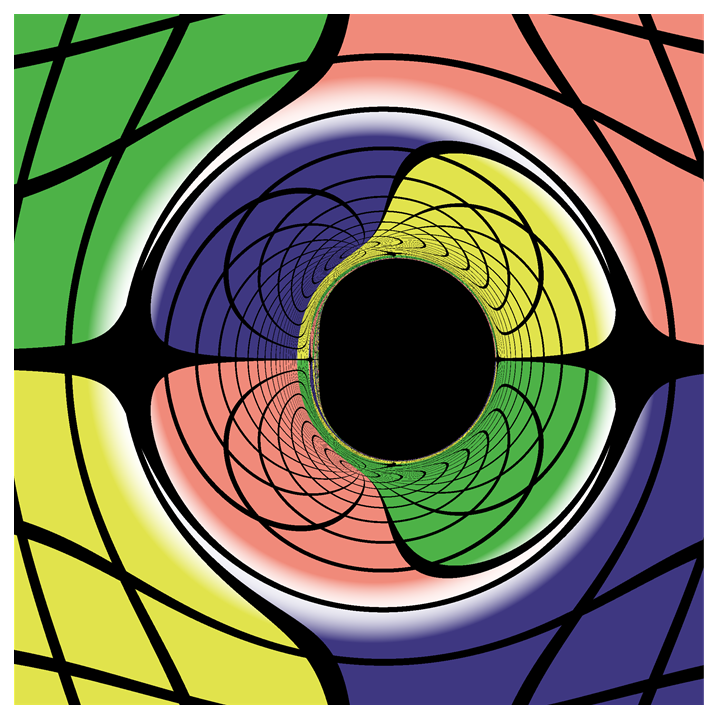}\label{fig:lensing2}}\hspace{1mm}
\subfigure[\, Kerr, $a=-0.999$]
{\includegraphics[width=4.cm]{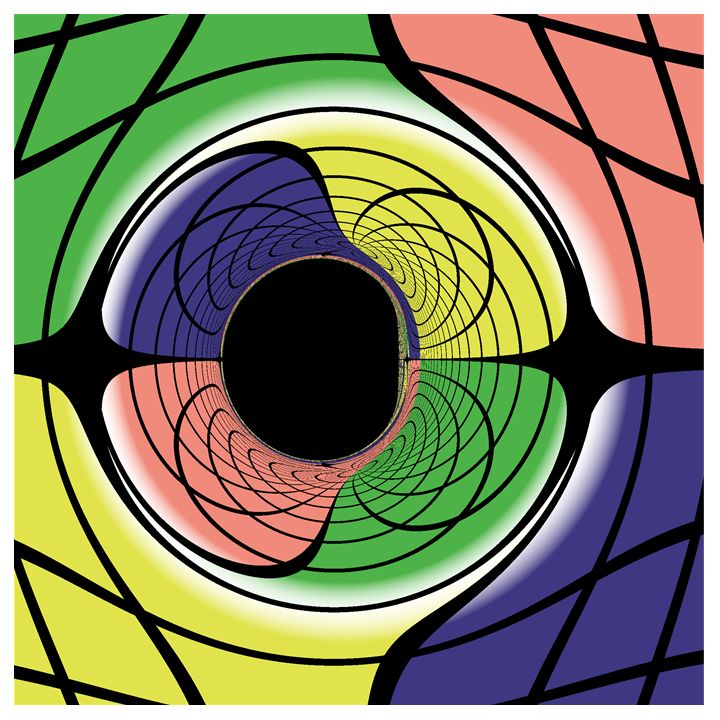}\label{fig:lensing3}}\\
\subfigure[\, $\gamma=0$, $a=0.2$]
{\includegraphics[width=4.cm]{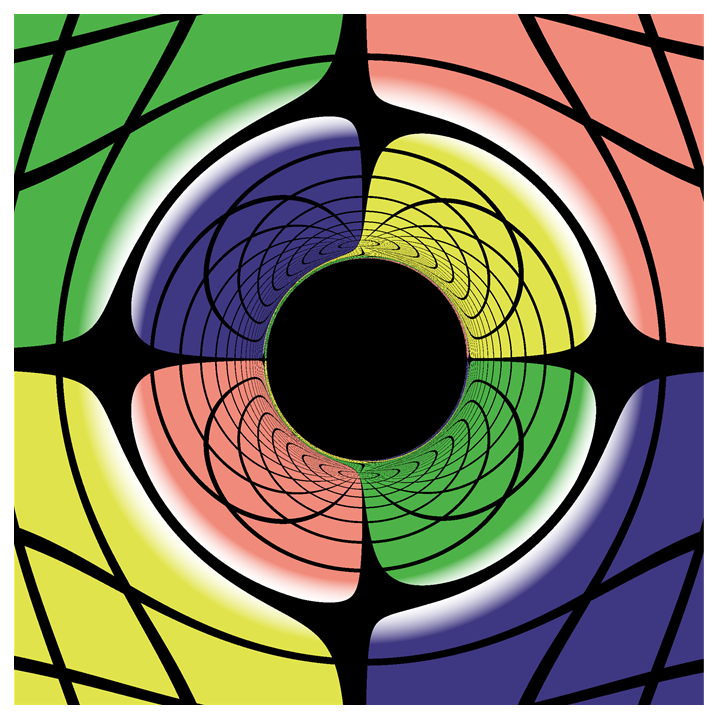}\label{fig:lensing4}}
\subfigure[\, $\gamma=1$, $a=0.2$]
{\includegraphics[width=4.cm]{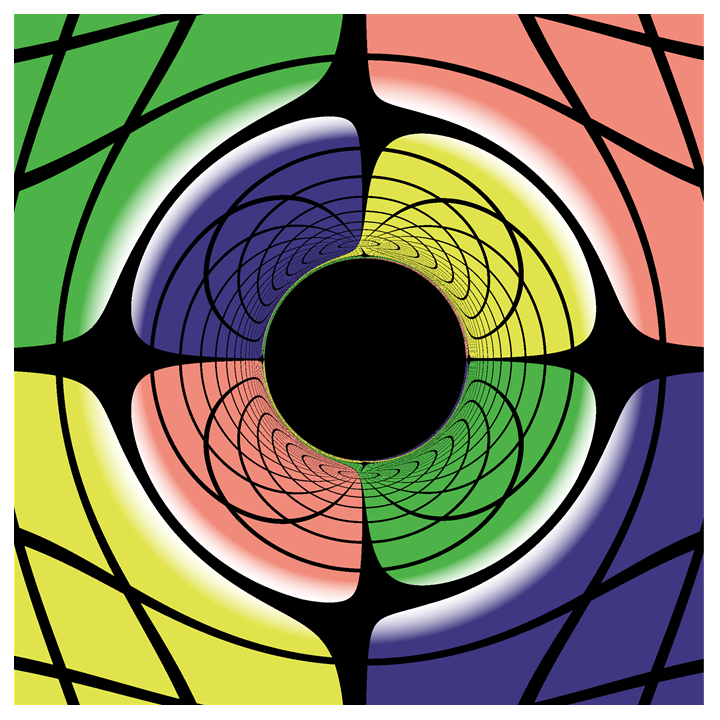}\label{fig:lensing5}}\hspace{1mm}
\subfigure[\, $\gamma=1.5$, $a=0.2$]
{\includegraphics[width=4.cm]{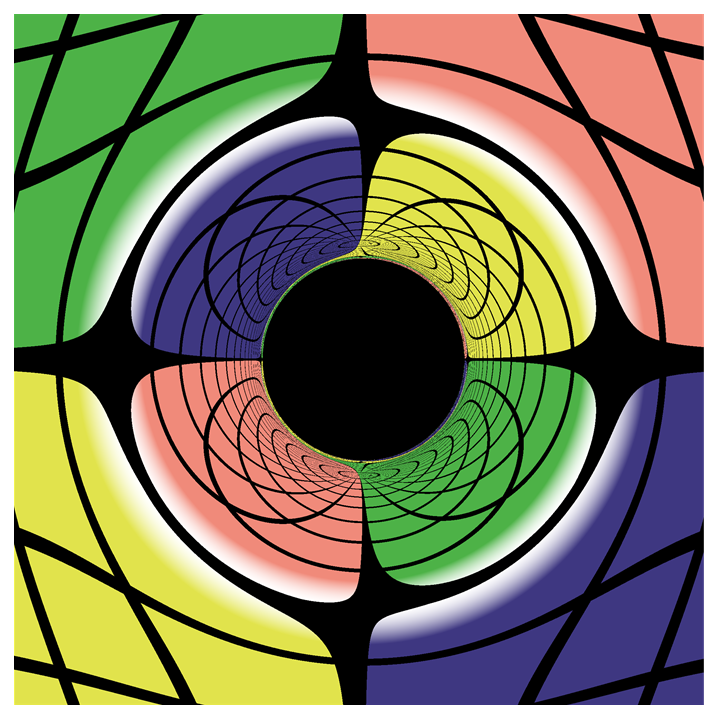}\label{fig:lensing6}}\hspace{1mm}
\subfigure[\, $\gamma=2$, $a=0.2$]
{\includegraphics[width=4.cm]{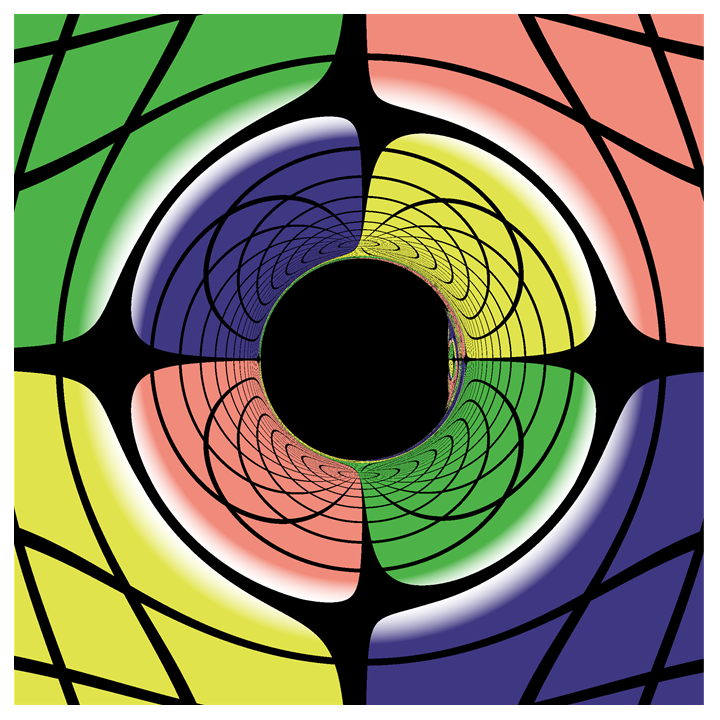}\label{fig:lensing7}}\hspace{1mm}
\caption{{Shadow and Lensing by Schwarzschild BH, Kerr BHs and slowly rotating Kerr-like BHs}. Here the inclination angle of the observer is fixed at $\theta_o=\pi/2$.}
\label{fig:lensing}
\end{figure}

As shown in  Fig. \ref{fig:source1}, the ball-like extended source is divided into four symmetric parts, each part is marked with a different color, and the intersection of the four colors is marked with a white dot. The black longitudinal and latitudinal lines divide the ball-like source into the grids, and the interval between adjacent longitudinal or latitudinal lines is $\pi/18$.
In order to investigate the trajectories of light rays, we can take the orthonormal tetrad at the observer's position
\begin{eqnarray}
\begin{split}
e_{0}=\frac{g_{\phi\phi}\partial_{t}-g_{\phi t}\partial_{\phi}}{\sqrt{g_{\phi\phi}(g^2_{\phi t}-g_{\phi\phi}g_{t t})}}, ~~~e_{1}=-\frac{\partial_{r}}{\sqrt{g_{rr}}},~~~e_{2}=\frac{\partial_{\theta}}{\sqrt{g_{\theta\theta}}},~~~e_{3}=-\frac{\partial_{\phi}}{\sqrt{g_{\phi\phi}}},
\label{eq-tetrad}
\end{split}
\end{eqnarray}
where $g_{\mu\nu}$ is the metric of Kerr-like BH. Since the optical paths are reversible, the light rays reaching the observer can be considered to be emitted by the observer. These light rays have two endings, either returning to the light source or falling into the BH, where the light rays falling into the {BH} correspond to the shadow of the BH.

In Fig. \ref{fig:lensing}, the images of Schwarzschild BH, Kerr BHs $(a=0.999, -0.999)$, and Kerr-like BHs with different coupling parameters $\gamma$ are plotted. The observer is placed at $\theta_o=\pi/2$. The black area in the center is the shadow of the BH {which has obvious symmetry}, and the white ring outside the shadow is the Einstein ring that can be clearly observed. The shadow of Schwarzschild {BH} is a perfect circle as can be seen in Fig. \ref{fig:lensing1}. Once we turn on the rotation in Kerr BH, $D$-shape will appear but it is faint, unless the spin $a$ is large enough, see for examples $a=0.999$~(Fig. \ref{fig:lensing2}) and $a=-0.999$~(Fig. \ref{fig:lensing3}).

From the figure it is easy to understand that for $a=0.2$ and  $\gamma=0$, a tiny distortion exists in the left side of the central shadow because of the co-rotating. As we increase $\gamma$, the left distortion will become fainter because increasing $\gamma$ makes the BH angular momentum decrease but not yet change the rotating direction.  There should be a critical $\gamma_c$ for which the angular momentum vanishes such that the shadow recovers perfect circle.  If we further increase $\gamma$ beyond $\gamma_c$, as shown in \cite{Chatzifotis:2022ene}   the BH angular momentum increases while it starts to rotate to the opposite direction, so the shadow distortion could appear in the other side because of the counter-rotating effect. This indeed explicitly happens in Fig. \ref{fig:lensing7} which is similar to that in Kerr BH with $a=-0.999$ in Fig. \ref{fig:lensing3}. Therefore, starting with a slowly co-rotating BH with $a=0.2$, as we increase the coupling of the axion field to the CS term, because there is an exchange of energy between the axion field and the strong gravity, we  observe a shadow which mimics that of a fast counter-rotating Kerr BH.


\begin{figure}[h]
\centering
\subfigure[\, $\gamma=2.5$, $a=0.2$]
{\includegraphics[width=4cm]{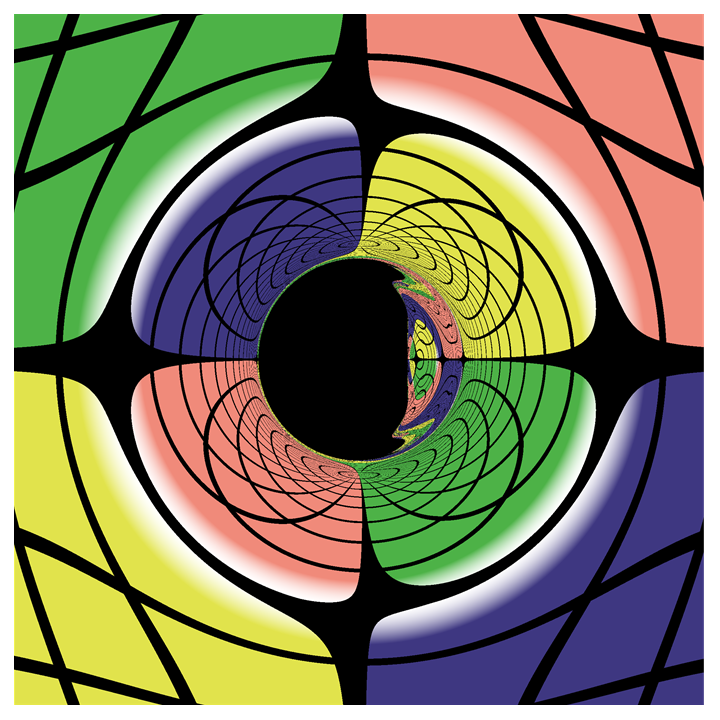}\label{fig:Images1}}\hspace{2mm}
\subfigure[\, $\gamma=3$, $a=0.2$]
{\includegraphics[width=4cm]{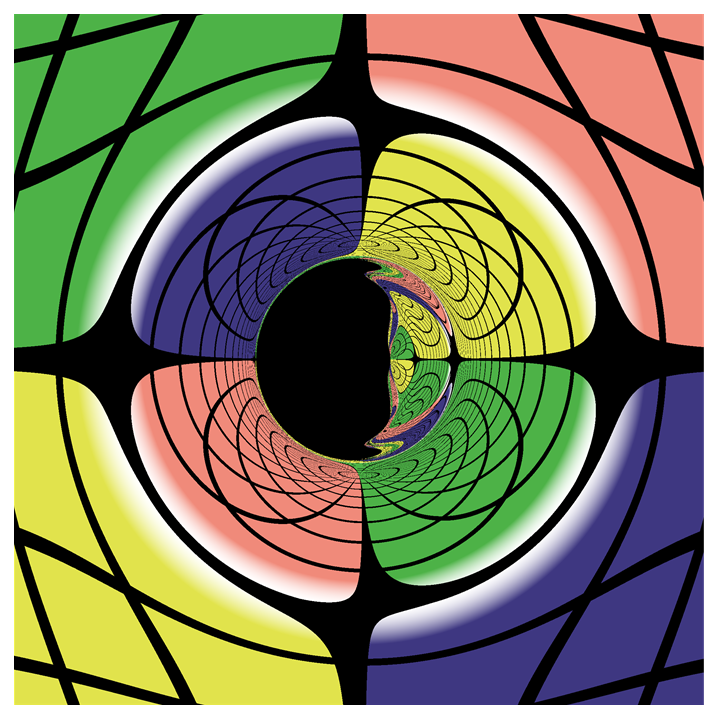}\label{fig:Images2}}\hspace{2mm}
\subfigure[\, $\gamma=4$, $a=0.2$]
{\includegraphics[width=4cm]{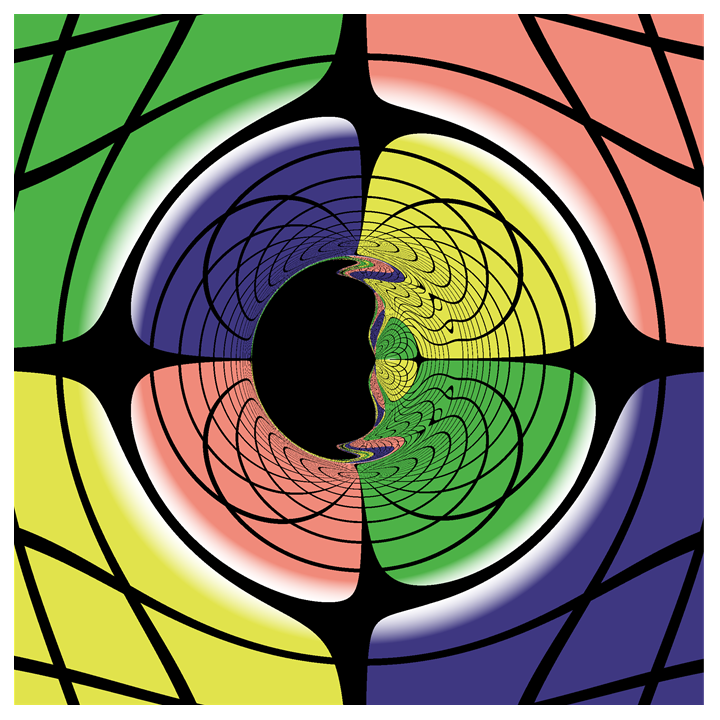}\label{fig:Images3}}\hspace{2mm}
\subfigure[\, $\gamma=5$, $a=0.2$]
{\includegraphics[width=4cm]{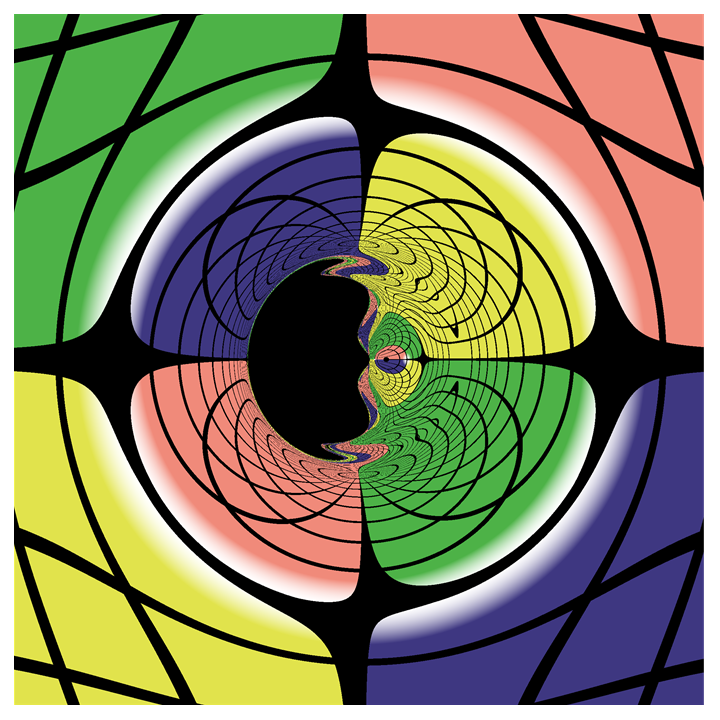}\label{fig:Images4}}
\caption{Shadow and Lensing by slowly  rotating Kerr-like BHs with large enough coupling parameters. Here the inclination angle of the observer is fixed at $\theta_o=\pi/2$.}
\label{fig:Images}
\end{figure}

\begin{figure}[h]
\centering
\subfigure[\, $\gamma=2.5$, $a=0.2$]
{\includegraphics[width=4cm]{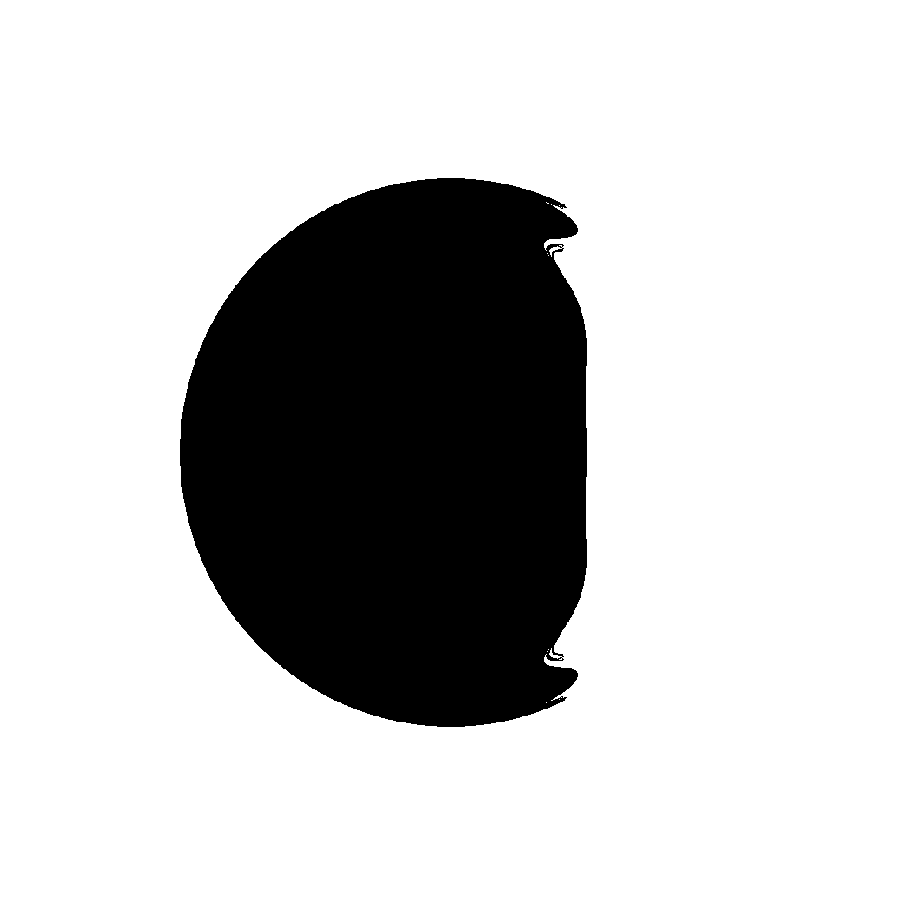}}\hspace{2mm}
\subfigure[\, $\gamma=3$, $a=0.2$]
{\includegraphics[width=4cm]{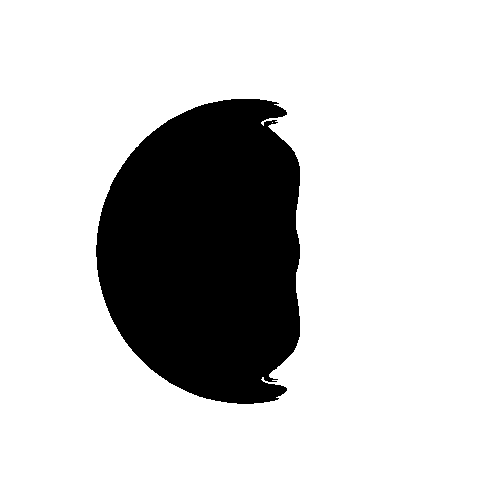}}\hspace{2mm}
\subfigure[\, $\gamma=4$, $a=0.2$]
{\includegraphics[width=4cm]{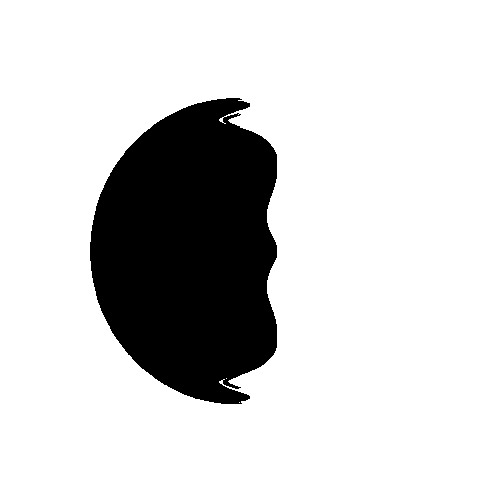}}\hspace{2mm}
\subfigure[\, $\gamma=5$, $a=0.2$]
{\includegraphics[width=4cm]{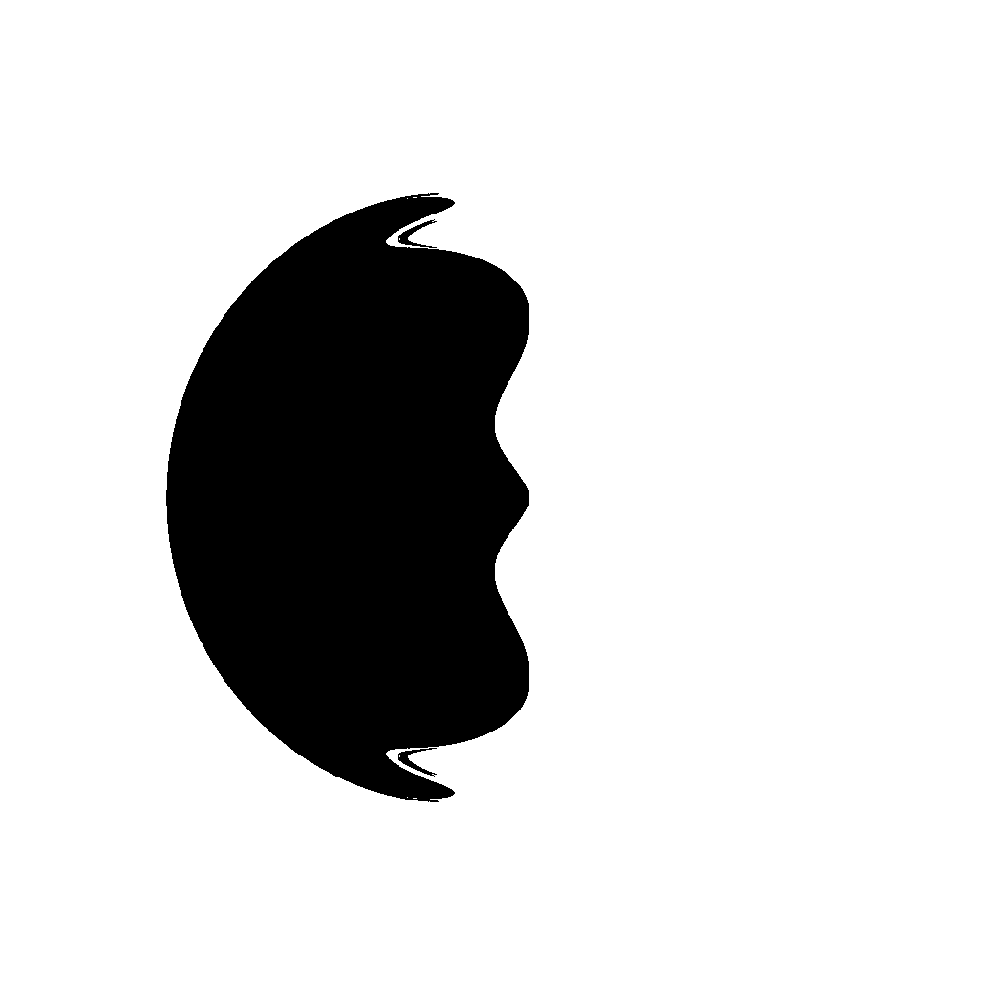}}
\caption{Shadow boundaries of the slowly  rotating Kerr-like BHs extracted from Fig. \ref{fig:Images}.}
\label{fig:shadow-g}
\end{figure}

In Fig. \ref{fig:Images} and Fig. \ref{fig:shadow-g}, the shadows and lensing {effects} of the slowly rotating Kerr-like BH with larger coupling parameters $\gamma$ are plotted. Comparing to the Kerr BH, we observe two novel structures of the shadow for the Kerr-like BH with $a=0.2$ : (i) The deformation of shadows of the slowly rotating Kerr-like BH with  positive $a=0.2$ is on the same side of Kerr BH with negative $a$. Moreover, though the whole spacetime rotates slowly, the distortion of the shadow is significant. This means that non-trivial interaction between the axion and gravity could construct a slowly rotating spacetime as a whole. But it has an internal structure in which a highly rotating event horizon is allowed, leading to highly rotating effects. (ii) Instead of standard $D$-shaped shadow for Kerr BH, here we observe ``human-face-like" shaped shadow. This may provide a new attempt for us to use very large baseline interferometry (VLBI) observations \cite{EventHorizonTelescope:2019ths,EventHorizonTelescope:2022apq} of the shadows to verify the existence of axion matter. Moreover, two disconnected crescent-shaped shadows appeared directly above and below the ``human-face-like" shaped shadow, which is more explicit in Fig. \ref{fig:shadow-g}. {It is noted that the critical shadow curves of other slowly rotating CS BHs have been studied in \cite{Amarilla:2010zq,Meng:2023wgi}, in which the novel shadow shapes in the current model were not observed. }
Besides, the lensing effects, in particular in the vicinity of the distorted shadow, are novel and different from Kerr BH. As we compare the lensing from Fig. \ref{fig:Images1} to Fig. \ref{fig:Images4}, we detect that a regime with pink and blue lights will merge on the symmetric axial near the distorted shadow curve, as the coupling $\gamma$ increases. It means that in this Kerr-like BH, some photons can have more than one radial turning point, generically corresponding to chaotic motion \cite{Zelenka:2017aqn,Destounis:2021mqv,Destounis:2020kss}, which distinguish from only one radial turning point in Kerr BH \cite{Wilkins:1972rs}. The disconnected shadows and lensing show that the interaction between the axionic matter and gravity, and the distribution of axion significantly affect the chaotic scattering, so does the chaotic region.


\section{Conclusion and discussion}
Starting from an axion modified Kerr-like BH, we observe that a slowly rotating Kerr-like BH can produce a similar $D$-shaped shadow as that for a rapidly counter-rotating Kerr BH. These phenomena seem to be a natural consequence of the energy exchange between the axion field and BH such that a slowly rotating spacetime can also form a rapidly rotating event horizon. As  the interplay becomes stronger,  novel ``human-face-like" shaped shadows and peculiar lensing phenomena, which significantly shift from those of Kerr BH, can be observed. Apparently, the rich lensing effects attribute to the chaotic scattering of the photons around the Kerr-like BH. An exhaustive studies on the bound orbits, light rings, and  trapped  trajectories \cite{Cunha:2016bjh} could help us to obtain a deeper understanding of all those phenomena, which shall be present elsewhere. In addition, it will be worthwhile to consider fast rotating Kerr-like BH to further study the effects among the fast rotation parameter $a$  and coupling parameter $\gamma$ on the shadows in the future.

\begin{acknowledgments}
This work is partly supported by the Natural Science Foundation of China under Grants No. 12375054,  Natural Science Foundation of Jiangsu Province under Grant No. BK20211601, and  the Postgraduate Research $\&$ Practice Innovation Program of Jiangsu Province under Grants No. KYCX22$_{-}$3452.
\end{acknowledgments}

\bibliographystyle{utphys}
\bibliography{ref}

\end{document}